# Wave vector substar group in reciprocal lattice space and its representation

**Il Hwan Kim[a], Jong Ok Pak[b], Il Hun Kim[a, c], Song Won Kim[a], Lin Li[c]**

[a] *Department of physics, Kim Hyong Jik Normal University, Pyongyang , Democratic People's Republic of Korea*
[b] *Department of physics, Pyongyang University of Mechanical engineering, Pyongyang , Democratic People's Republic of Korea*
[c] *College of Science, Northeastern University, Shenyang 110819, People's Republic of China*

Corresponding author: ririn@sohu.com (L. Li)

**Abstract** In the paper we establish the new conception of the wave vector substar group and its representation that, in the study on translational symmetry breaking of crystal, can only consider the particular arms of wave vector star taking part in phase transition, but not all arms of wave vector star, unlike the traditional Landau theory. We show that using the new conception, we can effectively investigate the interesting physical properties of crystals associated with translational symmetry breaking. We can see that the studies on the complicated phase transitions related to reducible representations, such as phase transitions in perovskite $KMnF_3$ multiferroics crystal and high temperature superconductor $La_{2/3}Mg_{1/2}W_{1/2}O_3$ ($La_4Mg_3W_3O_{18}$) compound, are much simplified by the new conception, and the theory of the wave vector substar group and its representation becomes a powerful mathematical tool being able to promote actively the study of various symmetry breaking phenomena of solid state crystals.

Keywords: wave vector substar group; representation of wave vector substar group; star channel group; translational symmetry breaking

## 1. Introduction

Application of the group theory in studying the physical properties of solid relates to the representation of group, because the physical properties of crystals are associated with the action of representation of space group(Landau & Lifshitz, 1980; Dresselhaus & Dresselhaus, 2008; Lyubarskiĭ, 2006). Already, previous researchers have called the set of elements of space group $G_0$ leaving the wave vector invariant the wave vector group. The full irreducible representations of space group are induced from the representation of the wave vector group.

All possible changes in symmetry of crystal which undergoes second-order phase transition were first studied by Landau in 1937(Landau & Lifshitz, 1980). Landau suggested the idea that the change of the density function of charge distribution in solid can be expanded in a complete set of the basis functions of the full irreducible representation of the space group $G$ of high-symmetry phase. The number of coefficients of the basis functions of irreducible representation inducing the phase transition, that is, the number of the order parameter components is equal to the dimension of irreducible representation. The dimension of representation relates with the number of arms of wave vector star taking part in the phase transition.

Many structural phase transitions occurring in crystals are just described by the full irreducible representations and the reducible representations of 230 space groups, and at the same time are related with one wave vector star or two and over wave vector stars. Generally, if the order of group is great and the dimension of representation is high, the research associated with the full irreducible representation is very complicated.

Recently, much attention has been focused on the phase transitions occurring in crystals such as multiferroics with interesting properties. For example, ferromagnetic(ferroelectric)-ferroelastic phase transitions of $RCrO_3$(R=Y, La, Pr, Sm, Dy, Ho, Yb, Lu) perovskite crystals(Weber *et al.*, 2012), antiferromagnetic-ferromagnetic phase transition of $KMnF_3$ crystals(Carpenter *et al.*, 2012) and antiferrodistortive phase transition of EuTiO$_3$ crystals(Goian *et al.*, 2012; Weber *et al.*, 2012) and so on are just occurred by reducible stars in reciprocal lattice space.

In Kim & Pak (2001), to overcome difficulties in studying the complicate phase transitions, the wave vector star channel group was defined, which is the set of elements of space group leaving the wave vector star channel(Naish & Syromyatnikov, 1976 ; Izyumov & Syromyatnikov, 1990) of brillouin zone in reciprocal lattice space invariant. Using this conception, in Kim & Pak (2007) and Ri *et al.* (2013), it showed that the symmetry breaking $O_h^1 \xrightarrow{319K} D_{4h}^5 \xrightarrow{317K} D_{2h}^{17} \xrightarrow{310K} D_{2h}^{16}(C_{2h}^2)$ in CsPbCl$_3$ ferroelastic crystal(Aleksandrov *et al.*, 1981; Gufan, 1982; Larin, 1984) could be described by the representation of the group of the star channel $[\boldsymbol{k}_{11}^{(i)} + \boldsymbol{k}_{13}](i=\overline{1,3})$ with the point group $D_{4h}$ as the parent phase symmetry, but no the full irreducible representations($M_3 \oplus R_{25}$ (Aleksandrov *et al.*, 1981; Gufan, 1982; Larin, 1984)) of the space group $O_h^1$. Also, in Kim & Jang (2010), it was shown that the phase transitions in Pb$_{1-x}$Ca$_x$TiO$_3$(PCT) crystal(Torgashev *et al.*, 2006(1); Torgashev *et al.*, 2006(2); Redfern, 1996) could be described by the group of the star channel $[\boldsymbol{k}_{11}^{(i)} + \boldsymbol{k}_{12} + \boldsymbol{k}_{13}](i=\overline{1,3})$ with the point group $D_{4h}$.

The results of Kim & Pak (2001), (2007), Ri *et al.* (2013) and Kim & Jang (2010) showed that, if we apply the conception of wave vector star channel group and its representation, we can use both the representation that its dimensional number is lowered and the integral rational basis of invariants that its number is very decreased, and, therefore, both symmetry research and phenomenological study can be very simplified.

Until now, in the field of studying on the symmetry of the structural phase transition, the Landau theory of second-order phase transition has been widely using, which is based on the full irreducible representation of the space group related with all arms of the wave vector star. If the Landau theory is based on the full irreducible representation of group of the wave vector substar related with the particular arms of wave vector star taking part in phase transition, it is no doubt to make greater advance in the study of phase transition. But, until now, there is no theory of the full irreducible representation of the group associated with the wave vector substar of the reciprocal lattice space. And we cannot discuss the all of the wave vector substars because the star channel was only defined about some of the wave vector stars(Jaric, 1983, Jaric & Senechal, 1984).

Now, there is no name of the group related to some arms(such as two arms, three arms, …) of wave vector star and, even, the name of the group related to the set of elements leaving all arms of star invariant, too. But, we have been widely using the representation of this group. Perhaps, in the viewpoint of previous researchers, the set of rotation symmetry elements leaving all arms of star invariant in the reciprocal lattice space is the set of elements of the real space, and thus, researchers did not feel the need to call the name of the group.

But, now, it is necessary for us to have a name of the group related to the substar of wave vector star in the reciprocal lattice space. That is why we can consider the translation symmetry breaking of crystal with the particular arms of the wave vector star.

The outline of this paper is as follows.

In Sec. 2, we are going to establish the theory of the wave vector substar group in the reciprocal lattice space. In Sec. 3, we are going to establish the irreducible representation theory of the substar group. In Sec. 4, we will show the applied examples of the representation theory of the substar group in contrast with the previous results.

## 2. The group of the wave vector substar in reciprocal lattice space

### 2.1. The wave vector group

The wave vector group is a subgroup of space group *G*. The wave vector group is formed by the set of space group elements $g=(h_i|\tau)$ which transform wave vector of reciprocal lattice space, $\boldsymbol{k}_\nu$, into itself, or into an equivalent vector $k_\nu = k_\nu + b$, where $b$ is the reciprocal lattice vector(Bradley & Cracknell, 1972; Kovalev & Hatch, 1993).

The wave vector group shows both the point symmetry and the translational symmetry of crystal. Obviously, $G_{k_\nu}$ is the space group of reciprocal lattice space. Here, we put special stress that the wave vector group $G_{k_\nu}$ is no the group related to symmetry operations in the reciprocal lattice space. The number of the wave vector groups related with 80 Lifshitz stars of 230 space groups is all 1350(Kovalev & Hatch, 1993).

**2.2. The wave vector substar group**

The set of space group elements which transform substar $[\cdots, k_i, \cdots, k_j, \cdots] = [k]$ of the star $\{k\}$ into itself or into equivalent vectors $[k] = [k] + b$, where $b$ is the reciprocal lattice vector, is formed the group. We call this the wave vector substar group and express to $G_{[k]}$.

$$G_{[k]} = \{g_n | g_n[k] = [k] + b\} \tag{1}$$

As it defines the wave vector group this way, it can be said that the substar group $G_{[k]}$ is the group which the wave vector group $G_k$ is generalized about the substar $[k]$, i.e., $G_k$ is a special case of $G_{[k]}$. Generally, the relation $G_k \not\subset G_{[k]}$ is established between $G_k$ and $G_{[k]}$. The group of the substar $G_{[k]} \equiv G_{\{k\}}$ related with all arms of the given wave vector star $\{k\}$ corresponds to the full symmetry space group $G$.

$G_{[k]}$ is the space group, too, and so, it contains the translation group $T_{[k]}$, infinite group, as a invariant subgroup. We call this the translation group of substar.

In case of Lifshitz stars among the wave vector stars $\{k\}$, the translation group $T_{[k]}$ is decomposed to cosets with respect to kernel $\ker \Gamma(t)$ of representation as follows:

$$T_{[k]} = (h_1|a_1)\ker\Gamma(t) + (h_1|a_2)\ker\Gamma(t) + \cdots + (h_1|a_s)\ker\Gamma(t), \tag{2}$$

where elements $(h_1|a_s)$ are coset representatives of the decomposition and $(h_1|a_1) = 1$.

From Eq. (2), we can define the factor group which the translation group $T_{[k]}$ is divided by the kernel $\ker\Gamma(t)$ of the representation. This group is just the finite translation group $\tilde{T}_{[k]}$, that is,

$$\tilde{T}_{[k]} = T_{[k]} \big/ \ker\Gamma(t). \tag{3}$$

Using Eq. (3), we can consider the finite translation group $\tilde{T}_{[k]}$ instead of infinite translation group $T_{[k]}$.

If $G_{[k]}$ is decomposed to cosets with respect to $T_{[k]}$, it is as follows:

$$G_{[k]} = g_1 T_{[k]} + g_2 T_{[k]} + \cdots + g_s T_{[k]}. \tag{4}$$

Because $T_{[k]}$ is a invariant subgroup of $G_{[k]}$, the factor group $G_{[k]} \big/ T_{[k]}$ can be defined. This factor group is isomorphic to the point group $G^0_{[k]}$.

The point group $G^0_{[k]}$ of substar of the given wave vector star coincides with one among 32 crystallographic point groups. The order of point group is $G = [G^0_{[k]}] = 1, 2, 3, 4, 6, 8, 12, 16, 24, 48$.

Because the wave vector substar group $G_{[\mathbf{k}]}$ is the space group too, it can be written as follows.

$$G_{[\mathbf{k}]} = G^0_{[\mathbf{k}]} \wedge T_{[\mathbf{k}]} \tag{5}$$

$G_{[k]}$ is an infinite group too, because $T_{[k]}$ is an infinite group. But in the case of Lifshitz star, we can consider

the finite group $\tilde{G}_{[\mathbf{k}]}$ instead of infinite group $G_{[\mathbf{k}]}$.

$$\tilde{G}_{[\mathbf{k}]} = G^0_{[\mathbf{k}]} \wedge \tilde{T}_{[\mathbf{k}]} \tag{6}$$

The order of finite translation group is $[\tilde{T}_{[\mathbf{k}]}]$=1, 2, 3, 4, 6, 8, 16, 32. Thus, the order of the finite space group $\tilde{G}_{[\mathbf{k}]}$ is $1 \leq [G^0_{[\mathbf{k}]}] \leq 1536$.

The number of Lifshitz wave vector substars in reciprocal lattice space is all 250(37 among 287 substars are not related with the symmetry breaking), and 179 substars among those are the wave vector star channels(Kim & Pak, 2001). There are 3649 wave vector substar groups related to 80 stars on 14 Bravais lattices of 7 crystal systems. Among those, there are 2811 wave vector star channel groups. For example, in the Appendix A we show Lifshitz wave vector substars, star channels and Lifshitz wave vector substar groups with respect to the simple cubic lattice($\Gamma_c$).

The wave vector group $G_{k_\nu}$ is the little group of the given wave vector star $\{k\}$. As the case of the wave vector group $G_{k_\nu}$ this way, we are going to define the little group of the wave vector substar. The set of space group elements which transform one arm $k_\nu$ of the substar $[k]$ into itself or into an equivalent vector $k_\nu = k_\nu + b$, where $b$ is the reciprocal lattice vector, forms a group. We call this the little group $G'_{k_\nu}$ of the substar $[k]$.

$$G'_{k_\nu} = \{g_i | g_i k_\nu = k_\nu + b\} \tag{7}$$

The denotation of $G'_{k_\nu}$ is in contrast with the little group $G_{k_\nu}$ of the wave vector star $\{k\}$. Generally, there is the relationship $G_{k_\nu} \supseteq G'_{k_\nu}$ between the little group $G_{k_\nu}$ of the wave vector star and the little group $G'_{k_\nu}$ of the substar.

Meanwhile, there is the relationship of the group-subgroup, $G_{[k]} \supseteq G'_{k_\nu}$, between the substar group $G_{[k]}$ and the little group $G'_{k_\nu}$ of substar. Therefore, the substar group $G_{[k]}$ can be decomposed to cosets with respect to $G'_{k_\nu}$:

$$G_{[k]} = g_1 G'_{k_\nu} + g_2 G'_{k_\nu} + \cdots + g_s G'_{k_\nu}, \tag{8}$$

where elements $g_i(i = \overline{1,s})$ are coset representatives of the decomposition and $g_i \notin G'_{k_\nu}$, $g_i \in G_{[k]}$. Elements $g_i(i = \overline{1,s})$ are the elements transforming arms of the given substar into one another. And the number *s* of coset representatives denotes the number of arms of substar.

If the little groups $G_{k_i}$ related to each arms of the substar are given, the substar group $G_{[k]}$ can be written as follows:

$$G_{[k]} = G_{[k_1, k_2, \cdots, k_n]} = \{G_{k_1} \cap G_{k_2} \cap \cdots \cap G_{k_n}\} \cup \{\{g_l\} \cap \{g_i\}\}, \tag{9}$$

where the first term $\{G_{k_1} \cap G_{k_2} \cap \cdots \cap G_{k_n}\}$ of Eq. (9) corresponds to a unit element of the permutation group $S_{k_n}$. $\{g_l\}$ of the second term $\{\{g_l\} \cap \{g_i\}\}$ corresponds to the $n!-1$ elements of the permutation group except a unit element, and $\{g_i\}$ contains the elements satisfying Eq. (8)($g_i \notin G'_{k_\nu}$, $g_i \in G_{[k]}$).

$$\begin{pmatrix} k_1, & k_2, & \cdots, & k_n \\ k_1, & k_2, & \cdots, & k_n \end{pmatrix} = (e) = \{G_{k_1} \cap G_{k_2} \cap \cdots \cap G_{k_n}\} \tag{10}$$

$$\begin{pmatrix} k_1, & k_2, & \cdots, & k_n \\ k_i, & k_j, & \cdots, & k_l \end{pmatrix} = (i, j, \cdots, l) = \begin{cases} g_l k_1 = k_i \\ g_l k_2 = k_j \\ \vdots \\ g_l k_n = k_l \end{cases} = \{\{g_l\} \cap \{g_i\}\} \quad (11)$$

Now, imagine of the substar $[k_1, k_2, \cdots, k_n]$.

From the viewpoint of the mathematics, n! permutation elements

$$\begin{pmatrix} k_1, & k_2, & \cdots, & k_n \\ P_{k_1}, & P_{k_2}, & \cdots, & P_{k_n} \end{pmatrix} = \begin{pmatrix} k_i \\ P_{k_i} \end{pmatrix} \quad (12)$$

form a group of the permutation $S_n(k_1, k_2, \cdots, k_n) \equiv S_{k_n}$. In Eq. (10), the group is formed by the set of the elements of space group corresponding to the unit element of the permutation group $S_{k_n}$. We express this as follows.

$$G_{k_1} \cap G_{k_2} \cap \cdots \cap G_{k_n} = G_e \quad (13)$$

$G_e$ is obviously an invariant subgroup. Therefore, the substar group $G_{[k]}$ is decomposed to cosets as follows.

$$G_{[k]} = g_1 G_e + g_2 G_e + \cdots + g_n G_e \quad (14)$$

From Eq. (14), we can define the factor group of the substar group $G_{[k]}$ with respect to $G_e$.

$$G_f = G_{[k]} / G_e \quad (15)$$

The factor group $G_f$ is generally isomorphic to the subgroup of n-dimensional permutation group $S_{k_n}$.

By the result that all the possible substar groups were found out, the groups related to some substars do not exist, for example, substar $[k_8^{(1234),(1256),(3456)}]$ of space groups $T^2, T_h^{3,4}$ of face–centered cubic lattice, substar $[k_{10}^{(12),(13),(23)}]$ of space groups $T^{1,4}, T_h^{1,2,6}$ of simple cubic lattice, substar $[k_{11}^{(12),(14),(23),(34)}]$ of space groups $C_4^{5,6}, S_4^2, C_{4h}^{5,6}$ of body–centered tetragonal lattice, substar $[k_4^{(12),(13),(23)}]$ of space groups $C_3^4, S_6^2$ of rhombohedral lattice, substar $[k_{14}^{(12),(13),(23)}]$ of space groups $C_{6h}^{1,2}, C_{3h}^1, C_3^{1-3}, S_6^1, C_6^{1-6}$ of hexagonal lattice.

### 2.3. The wave vector star group

As above mentioned, the wave vector group is the set of elements of space group $G$ leaving one arm of the given wave vector star $\{k\}$ invariant and the wave vector substar group is the set of elements of space group $G$ leaving one substar $[k_1, k_2, \cdots, k_n]$ of the given star invariant.

From this viewpoint, we are going to name the group $G_{\{k\}}$ the wave vector star group. The wave vector star group is the set of the elements of space group $G_{\{k\}}$ leaving all arms of the star $\{k\}$ invariant. The group is expressed as follows.

$$G_{\{k\}} = \{g_n | g_n \{k\} = \{k\} + b\} \quad (16)$$

To leave the wave vector star invariant means that this group contains all elements transforming arms of star into one another.

The wave vector group is the subgroup of the wave vector star group. We call usually this group the little group.

Thus, the wave vector star group $G_{\{k\}}$ can be decomposed to cosets with respected to the wave vector group $G_{k_v}$, little group: ,

$$G_{\{k\}} = G_{k_v} + g_2 G_{k_v} + \cdots + g_L G_{k_v} = \sum_{i=1}^{L} g_i G_{k_v}, \qquad (17)$$

where elements $g_i$ is coset representatives of decomposition and $g_i \in G_{\{k\}}$.

From Eq. (17), the representation of the wave vector star group is induced from the representation of the little group. This means that to find out the full irreducible representations of 230 space groups is just equal to find out the full irreducible representations of the wave vector star groups.

Because the wave vector group is related to one arm of the star, the numbers of the wave vector groups and the wave vector star groups related to all arms of star are equal.

In the case of Lifshitz stars, the wave vector star groups is groups having 1, 2, 3, 4 and 6 arms, respectively.

If we only consider the Lifshitz stars of Brillouin zone, we have 80 stars, 1350 wave vector groups and 1350 wave vector star groups.(Table 1)

**Table 1**. The number of Lifshitz substar groups of 230 space groups

| Lattice Type | Space Group | Substar Group | | | | | | Star Channel Group | Little group | Star Group |
|---|---|---|---|---|---|---|---|---|---|---|
| | | 1 arm | 2 arms | 3 arms | 4 arms | 6 arms | sum | | | |
| $\Gamma_t$ | 2 | 16 | | | | | 16 | 16 | 16 | 16 |
| $\Gamma_m$ | 8 | 64 | | | | | 64 | 64 | 64 | 64 |
| $\Gamma_m^b$ | 5 | 40 | 10 | | | | 50 | 50 | 30 | 30 |
| $\Gamma_o$ | 30 | 240 | | | | | 240 | 240 | 240 | 240 |
| $\Gamma_o^b$ | 15 | 120 | 30 | | | | 150 | 150 | 90 | 90 |
| $\Gamma_o^f$ | 5 | 40 | 30 | | 5 | | 75 | 75 | 25 | 25 |
| $\Gamma_o^v$ | 9 | 90 | 36 | | | | 126 | 108 | 54 | 54 |
| $\Gamma_q$ | 49 | 392 | 98 | | | | 490 | 490 | 294 | 294 |
| $\Gamma_q^v$ | 19 | 190 | 132 | | 19 | | 341 | 303 | 95 | 95 |
| $\Gamma_{rh}$ | 7 | 56 | 30 | 14 | | | 100 | 85 | 28 | 28 |
| $\Gamma_h$ | 45 | 540 | 282 | 90 | | | 912 | 636 | 270 | 270 |
| $\Gamma_c$ | 15 | 120 | 60 | 30 | | | 210 | 180 | 60 | 60 |
| $\Gamma_c^v$ | 10 | 100 | 112 | 64 | 18 | 10 | 304 | 170 | 40 | 40 |
| $\Gamma_c^f$ | 11 | 154 | 219 | 143 | 44 | 11 | 571 | 244 | 44 | 44 |
| sum | 230 | 1840 | 706 | 185 | 59 | 21 | 3649 | 2811 | 1350 | 1350 |

**3. The irreducible representation of the wave vector substar group**

Generally, the method of little group has being widely used to find out the irreducible representation of group having the invariant subgroup(Lyubarskiĭ, 2006). The space group of crystal, $G$, includes the translation group $T$ as the invariant subgroup and, its irreducible representation can be found out by using the method of little group. In this method, the wave vector group $G_k$ is used as the little group and the full irreducible representation of space group $G$ related with all arms of star is induced by the irreducible representation of $G_k$ ( that is, the small representation).

The wave vector substar group $G_{[k]}$ is also the space group and contains the translation group as the invariant subgroup, and hence, its full irreducible representation can also be induced by the irreducible representation of its little group. Therefore, we should first of all find out the this representation.

Let us imagine of the wave vector substar $[k]$. And then, if the irreducible representation $\Gamma'_k$ of the little group $G'_k$ of the substar is known, we can induce the full irreducible representation $\Gamma_{[k]}$ of the substar group $G_{[k]}$ by Eq. (8). The irreducible representation $\Gamma'_k$ of the little group $G'_k$ of substar can be found out by applying the same method as it find out the irreducible representation of the wave vector group.

### 3.1. The irreducible representation of the wave vector group

The irreducible representation of the wave vector group $G_k$ is often called the small representation. Obviously, this representation is characterized by one arm of wave vector star(Mainly it means the first arm of star.). The small representation is expressed to $\Gamma_k$.

The method determining the small representation $\Gamma_k$ of $G_k$ is as follows.

We can separate the zero-block having elements of $(h|\tau_h)$-form and other blocks having elements of $(h|\tau_h + t_i)$-form from the wave vector group $G_k$ of any space group. The matrix of element $(h|\tau_h + t_i)$ for the i-th block differs by the exponent $\exp(-ikt_i)$ from one for the zero-block, as follows:

$$\Gamma_k((h|\tau_h + t_i)) = e^{-ik \cdot t_i} \cdot \Gamma_k((h|\tau_h)), \qquad (18)$$

where $\tau_h$ is improper translation element and $t_i$ is proper translation element. After all, if the matrices for zero-blocks of $G_k$ are only found out, we can determine the matrices of small representation $\Gamma_k$. The number of elements of zero-block is equal to the order of point group $G_k^0$. The set of all elements of zero-block does not form a group for itself, but the elements are expressed by the matrices of the projective representation $\Gamma_k^{\text{pro}}(h)$ of point group $G_k^0$.

In the handbook for the irreducible representation of space group(Kovalev & Hatch, 1993), there are tables of all the projective representations $\Gamma_k^{\text{pro}}(h)$. Therefore, we can easily find out the small representation $\Gamma_k$, that is,

$$\Gamma_k(g) = \Gamma_k^{\text{pro}}(h) \cdot exp(-ik\tau_h), \qquad (19)$$

where $g = (h|\tau_h)$ is the element of zero-block of the wave vector group.

### 3.2. The small representation of the wave vector substar group

Now, imagine of the wave vector substar $[k]$. Let us divide the cases of finding out the small representation into according to the relationship between the little group of substar, $G'_k$, and the wave vector group $G_k$.

#### 3.2.1. In the case of $G_{[k]}^0 \supset G_k^0$

In this case,

$$G_k^{\prime 0} = G_k^0. \qquad (20)$$

- In the case of $k_1 \in [k]$

The elements of the little group of substar are equal to the elements of the wave vector group, so that the small representation of substar group is equal to the small representation of space group.(where, $k_1$ - the first arm of wave vector star and also the first arm of wave vector substar.)

- In the case of $k_1 \notin [k]$

Between the elements of the wave vector group and the ones of the little group of substar, there is a conjugate relationship, and therefore, the small representation $\Gamma'_k$ of substar group and the one $\Gamma_k$ of space group conjugate each other. In the case, the small representation $\Gamma'_k$ of the substar group is found out using the small representation $\Gamma_k$ of space group and the following relation:

$$\Gamma'_k(g) = \Gamma_k(g_v^{-1} g g_v),  \qquad (21)$$

where the element $g_v$ transforms the first arm of substar into other arm($g_v k_1 = k_v$), $g_v^{-1} g g_v$ is the element of the wave vector group and $g$ is the element of substar group: $g_v^{-1} g g_v \in G_k$, $g \in G'_k$.

### 3.2.2. In the case of $G^0_{[k]} = G^0_k$

Between the little group of wave vector substar and the wave vector group there is the relationship;

$$G'_k \subset G_k. \qquad (22)$$

In the case, there is no the projective representation associated with the little group of wave vector substar in Kovalev & Hatch (1993). Hence, in order to determine the small representation $\Gamma'_k$ of substar group, we should newly find out the projective representation of this group by using the method of Bradley & Cracknell (1972) (so-called the method of factor group) or the one of Lyubarskiĭ (2006)(so-called the method of projective representation).

The formula to find out the projective representation is as follows(Lyubarskiĭ, 2006):

$$\hat{\Gamma}(h_i)\hat{\Gamma}(h_j) = \hat{\Gamma}(h_i h_j)\varphi(h_i h_j), \qquad (23)$$

$$\varphi(h_i h_j) = \exp[i(h_i^{-1} k - k, \tau_j)], \qquad (24)$$

where $\varphi(h_i h_j)$ is the exponential factor, $h_i, h_j$ is the rotational elements, $\tau_j$ is improper translational element corresponding to the rotational element $h_j$ and $k$ is the arm of the given wave vector substar.

### 3.3. The full irreducible representation of the wave vector substar group

**3.3.1. Method finding out the full irreducible representation $\Gamma_{[k]}$ of $G_{[k]}$ by the little group of substar $G'_k$**

Let us imagine of the substar $[k_1, k_2, k_3, \cdots, k_v]$.

Now, think of any element $\forall g(\in G_{[k]}, \notin G'_k)$, which transforms $k_1$ into $k_v$ by orthogonal transformation $h$.

$$h k_1 = k_v, \quad g = (h|\tau) \qquad (25)$$

According to Eq. (25), select the elements $g_2$, $g_3$, … and $g_v$ which are $g_2 k_1 = k_2$, $g_3 k_1 = k_3$, … and $g_v k_1 = k_v$. These elements are the cosets representatives of Eq. (8) satisfying $g_2, g_3, \cdots, g_v \in G_{[k]}$, $g_2, g_3, \cdots, g_v \notin G'_k$.

Now, let us the element $g$ ($g \in G$) act on the bloch function $\psi^{k_v}$, and then, the basis functions corresponding to arms of substar are transformed into one another. The full irreducible representation of the substar group $G_{[k]}$ just consists of the transformation matrices $\Gamma_{[k]}(g)$. The form of matrices can be written as follows.

$$\Gamma_{[k]}(g) = \begin{pmatrix} \Gamma'_k(g_1^{-1}gg_1) & \Gamma'_k(g_1^{-1}gg_2) & \cdots & \Gamma'_k(g_1^{-1}gg_v) \\ \Gamma'_k(g_2^{-1}gg_1) & \Gamma'_k(g_2^{-1}gg_2) & \cdots & \Gamma'_k(g_2^{-1}gg_v) \\ \vdots & \vdots & \vdots & \vdots \\ \Gamma'_k(g_v^{-1}gg_1) & \Gamma'_k(g_v^{-1}gg_2) & \cdots & \Gamma'_k(g_v^{-1}gg_v) \end{pmatrix} \quad (26)$$

Among elements of matrix in Eq. (26), there are also elements corresponding to $g_i^{-1}gg_i \notin G'_k$, $i = \overline{1,v}$. We denote the number 0 to places of these elements. Then, Eq. (26), is expressed by the elements of $G'_k$. After all, the full irreducible representation is induced by the representation of the little group of substar.

### 3.3.2. Method finding out the full irreducible representation $\Gamma_{[k]}$ of $G_{[k]}$ by the class

The wave vector substars are related to one another by symmetry, and so, we can consider those entirely in the way in which divide all substars into several classes. Because the substar groups belonging to the same class are conjugated one another, the representations related with those become the equivalent representations.

Now, like the case of the wave vector group, expand the space group $G$ into cosets of the subgroup $G_{[k]}$:

$$G = G_{[k]} + g_2 G_{[k]} + \cdots + g_N G_{[k]} = \sum_{i=1}^{N} g_i G_{[k]}, \quad (27)$$

where $g_i$-representative of coset and $g_i \notin G_{[k]}$, $g_i \in G$.

Clearly, the elements of the first coset leave the substar $[k]$ invariant and elements of other coset containing the representative $g_i$ transform the substar $[k]$ into different nonequivalent substar $[k]'$. Using all other representatives, we can find out all nonequivalent substars $[k]_i$.

The number N of nonequivalent substars is equal to the number of cosets. The nonequivalent substars can be found out by acting representatives on the first substar.

$$[k]_i = g_i [k] \quad (28)$$

Actually, because the rotational elements of space group act on the substar $[k]$, Eq. (28) can be rewritten as follows:

$$G^0 = G^0_{[k]} + h_2 G^0_{[k]} + \cdots + h_N G^0_{[k]} = \sum_{i=1}^{N} h_i G^0_{[k]}, \quad (29)$$

where $h_i$-the rotation factor of the element $g_i = (h_i | \tau_i)$ of space group.

The substars $[k]'_i$ expressed to Eq. (29) are transformed into one another under the action of elements of space group. Further, all of nonequivalent wave vector substars, which are obtained by acting all rotation elements of the group $G$ on the substar $[k]$, consist of the class of substar. Thus, wave vector substars are divided into some classes of substar.

The substar groups belonging to a class of substar conjugate one another.

$$G_{[k]'_i} = g_i G_{[k]} g_i^{-1} \quad (30)$$

Therefore, we can find out the irreducible representations of substar groups belonging to the same class as follows: First, it finds out the irreducible representation of any substar(which contains the first arm of the given star). Next, we can find out representations of other substars in the way in which obtain representations equivalent to it.

We can say that the substar group $G_{[k]}$ is the "little group" of the class $\{[k]\}$ of the given substar as the wave vector group $G_k$ is the little group of the given star $\{k\}$. The representation $\Gamma_{[k]_i}(g^{[k]_i})$ of the $G_{[k]'_i}$ is obtained by using the relationship

$$\Gamma_{[k]_i}(g^{[k]_i}) = \Gamma_{[k]}(g_i^{-1} g^{[k]_i} g_i), \tag{31}$$

where $g_i^{-1} g^{[k]_i} g_i = g^{[k]} \in G_{[k]}$.

Using the above method, we obtained the full irreducible representations related with 80 Lifshitz irreducible star channels of 14 Bravais lattices(Kim & Pak, 2016). For example, we show the full irreducible representations of Lifshitz wave vector star channel groups with respect to simple cubic lattice $O_h^1$ in the Appendix B.

**4. Comparison with previous results**

The full irreducible representation of the star channel group, its formation being very simple, has the far lower dimension than that related with all arms of star. Therefore, if the representation of substar group was used in studying the translational symmetry breaking of phase transition, it could be quite effective.

In Carpenter et al. (2012), the ferroelastic-ferromagnetic phase transition $O_h^1 \xrightarrow{185K} D_{4h}^{18} \xrightarrow{87K} D_{2h}^{17} \xrightarrow{82K} D_{2h}^{16}$ in perovskite KMnF$_3$ multiferroics crystal was investigated. By the previous results, this phase transition is described by the 9-component order parameter which is transformed according to the reducible representation related with $M$, $R$, $\Gamma$ points of Brillouin zone of space group $O_h^1$.

Reducible star $\{M, R, \Gamma\}$ consists of 5 arms $\{\frac{1}{2}(b_1 + b_2), \frac{1}{2}(b_1 + b_3), \frac{1}{2}(b_2 + b_3), \frac{1}{2}(b_1 + b_2 + b_3), 0\}$. Using the method of Kim & Pak (2001), let us find out the star channels, that is, the admitable substars related to the translational symmetry breaking. We accepted the space of 5-dimensional order parameter in which the representation of translation group acts on. And we determined the invariant vectors by finding out the subspaces invariant by the translation $L$-group. As a result, the star channels are [$\Gamma$], [$M_1\Gamma$], [$M_2\Gamma$], [$M_3\Gamma$], [$R\Gamma$], [$M_1M_2M_3\Gamma$], [$M_1R\Gamma$], [$M_2R\Gamma$], [$M_3R\Gamma$]. From the group-theoretical analysis, the phase transition is induced by the 5-dimensional reducible representation of the star channel group related with the star channel [$M_1R\Gamma$]. The 5-dimensional reducible representation can be obtained from the irreducible representations(Kim & Pak (2016)) $\tau(M)$, $\tau(R)$, $\tau(\Gamma)$ of the star channel groups $O_h^1[M^{(1)}]$, $O_h^1[R]$, $O_h^1[\Gamma]$.(Table 2)

**Table 2.** Representation of the star channel group $O_h^1[M^{(1)}R\Gamma]$

| Rep. \ elem. | $(h_2\|0)$ | $(h_3\|0)$ | $(h_{13}\|0)$ | $(h_{25}\|0)$ | $(h_1\|a_1)$ | $(h_1\|a_3)$ |
|---|---|---|---|---|---|---|
| $\tilde{T}$ | $\begin{pmatrix}1\\-1\\-1\\1\\1\end{pmatrix}$ | $\begin{pmatrix}-1\\1\\-1\\1\\-1\end{pmatrix}$ | $\begin{pmatrix}&&1&&\\1&&&&\\&&&1&\\&-1&&&\\&&&&-1\end{pmatrix}$ | $\begin{pmatrix}-1\\-1\\-1\\1\\-1\end{pmatrix}$ | $\begin{pmatrix}-1\\-1\\-1\\-1\\1\end{pmatrix}$ | $\begin{pmatrix}-1\\-1\\-1\\1\\1\end{pmatrix}$ |
| $\tau_1$ | $\begin{pmatrix}1\\-1\end{pmatrix}$ | $\begin{pmatrix}-1\\1\end{pmatrix}$ | $\begin{pmatrix}&1\\1&\end{pmatrix}$ | $\begin{pmatrix}-1\\-1\end{pmatrix}$ | $\begin{pmatrix}-1\\-1\end{pmatrix}$ | $\begin{pmatrix}-1\\-1\end{pmatrix}$ |
| $\tau_2$ | (-1) | (-1) | (1) | (-1) | (-1) | (-1) |
| $\tau_3$ | (1) | (1) | (-1) | (1) | (-1) | (1) |
| $\tau_4$ | (-1) | (-1) | (-1) | (-1) | (1) | (1) |

In Table 2, $\tau_1 \sim \tau_4$ are the irreducible representations of the star channel group $O_h^1[M^{(1)}R\Gamma]$ and $\tilde{T}$ is the reducible representation.(generating elements are only shown.) The expression of group elements follows to Kovalev & Hatch (1993).

As shown in Table 2, the star channel group $O_h^1[M^{(1)}R\Gamma]$ has the point group $D_{4h}$ and the dimension of representation is 5. Thus, the representation of this group is much simpler than the one of reducible star $\{M, R, \Gamma\}$ of space group $O_h^1$. If the representation of this group is used, the above phase transition is described as
$O_h^1[M^{(1)}R\Gamma] \xrightarrow{185K} D_{4h}^{18} \xrightarrow{87K} D_{2h}^{17} \xrightarrow{82K} D_{2h}^{16}$.

According to the method of Kim & Pak (2006), we analyzed the symmetry of the following reducible representations:

$$\tau_1 \oplus \tau_2, \ \tau_1 \oplus \tau_3, \ \tau_1 \oplus \tau_4, \ \tau_2 \oplus \tau_3, \ \tau_2 \oplus \tau_4, \ \tau_3 \oplus \tau_4, \ \tau_1 \oplus \tau_2 \oplus \tau_3,$$
$$\tau_1 \oplus \tau_2 \oplus \tau_4, \ \tau_1 \oplus \tau_3 \oplus \tau_4, \ \tau_2 \oplus \tau_3 \oplus \tau_4, \ \tilde{T} = \tau_1 \oplus \tau_2 \oplus \tau_3 \oplus \tau_4$$

As a result, the phase transition is described by the 5-componential order parameter.(Table 3)

**Table 3**. List of symmetry changes described by the star channel group $O_h^1[M^{(1)}R\Gamma]$

| Order parameter ($\vec{\eta}$) | Space group ($G_\alpha$) | Volume change ($V'/V$) |
|---|---|---|
| $(0\ 0\ 0\ \eta_4\ 0)$ | $D_{4h}^5$ | 2 |
| $(0\ 0\ \eta_3\ 0\ 0)$ | $D_{4h}^{18}$ | 2 |
| $(\eta_1\ \text{-}\eta_1\ 0\ 0\ 0)$ | $D_{2h}^{28}$ | 2 |
| $(\eta_1\ 0\ 0\ 0\ 0)$ | $D_{2h}^{23}$ | 1 |
| $(\eta_1\ 0\ 0\ \eta_4\ 0)$ | $D_{2h}^{17}$ | 4 |
| $(\eta_1\ \text{-}\eta_1\ 0\ \eta_4\ 0)$ | $D_{2h}^{16}$ | 4 |
| $(0\ 0\ \eta_3\ \eta_4\ 0)$ | $D_{4h}^5$ | 4 |

From Table 3, we cam see that the decrease of the symmetry of order parameter makes it study the phenomenological theory simply, so that we can consider the high-degree nonlinearity by it.

For phenomenological study, we must construct the thermodynamic potential.

According to the method of Sergienko & Gufan (2002), the integral rational basis of invariants(IRBI) of $G_{[M^{(1)}R\Gamma]} \equiv O_h^1[M^{(1)}R\Gamma]$ is as follows.

$$I_1 = \eta_1^2 + \eta_2^2, \ I_2 = \eta_1^2\eta_2^2, \ I_3 = \eta_3^2, \ I_4 = \eta_4^2, \ I_5 = \eta_5^2 \quad (32)$$

As shown Eq. (32), there is the basis with only one 4th invariant. But according to the previous method, there is the basis with 7 4th invariants(Weber *et al.*, 2012).

The 4th thermodynamic potential of 5-componential order parameter is as follows.

$$\Phi = a_1(\eta_1^2 + \eta_2^2) + a_2\eta_3^2 + a_3\eta_4^2 + a_4\eta_5^2 + b_1(\eta_1^2 + \eta_2^2)^2 + b_2\eta_1^2\eta_2^2 + b_3\eta_3^4 + b_4\eta_4^4 + b_5\eta_5^4 \quad (33)$$

The 4th thermodynamic potential of 9-componential order parameter is as follows.

$$\begin{aligned}
\Phi = &\, a_1(\eta_1^2 + \eta_2^2 + \eta_3^2) + a_2(\eta_4^2 + \eta_5^2 + \eta_6^2) + a_3(\eta_7^2 + \eta_8^2 + \eta_9^2) + b_1(\eta_1^2 + \eta_2^2 + \eta_3^2)^2 + \\
&\, b_2(\eta_4^2 + \eta_5^2 + \eta_6^2)^2 + b_3(\eta_7^2 + \eta_8^2 + \eta_9^2)^2 + b_4(\eta_1^2\eta_2^2 + \eta_2^2\eta_3^2 + \eta_1^2\eta_3^2) + b_5(\eta_1^2\eta_6^2 + \\
&\, \eta_2^2\eta_5^2 + \eta_3^2\eta_4^2) + b_6(\eta_1^2\eta_9^2 + \eta_2^2\eta_8^2 + \eta_3^2\eta_7^2) + b_7(\eta_4^2\eta_5^2 + \eta_4^2\eta_6^2 + \eta_5^2\eta_6^2) + b_8(\eta_4^2\eta_8^2 + \\
&\, \eta_4^2\eta_9^2 + \eta_5^2\eta_7^2 + \eta_5^2\eta_9^2 + \eta_6^2\eta_7^2 + \eta_6^2\eta_8^2) + b_9(\eta_4\eta_5\eta_7\eta_8 + \eta_4\eta_6\eta_7\eta_9 + \eta_5\eta_6\eta_8\eta_9) + \\
&\, b_{10}(\eta_7^2\eta_8^2 + \eta_7^2\eta_9^2 + \eta_8^2\eta_9^2) + b_{11}(\eta_1^2 + \eta_2^2 + \eta_3^2)(\eta_4^2 + \eta_5^2 + \eta_6^2) + b_{12}(\eta_1^2 + \eta_2^2 + \eta_3^2) \\
&\, (\eta_7^2 + \eta_8^2 + \eta_9^2) + b_{13}(\eta_4^2 + \eta_5^2 + \eta_6^2)(\eta_7^2 + \eta_8^2 + \eta_9^2) \cdots
\end{aligned} \qquad (34)$$

From Eqs. (33) and (34), it is clear that the symmetry breaking of perovskite KMnF$_3$ crystal can be effectively described by the simplified 5-componential order parameter model.

In Table 4, the Landau symmetry phases $G_\alpha$ related with Lifshitz star $\boldsymbol{k}_{10}$ of space group $O_h^2$ are showed in comparison with the result of Stokes & Hatch (1988).

In Stokes & Hatch (1988) the Landau symmetry phases were induced by the full irreducible representation of space group related with all arms of star. As shown, the phases induced by the full irreducible representation of substar group approximately coincide with the result of Stokes & Hatch (1988).

In Table 4, $\boldsymbol{k}_{10}^{(1)}$ denotes the first arm of Lifshitz star $\boldsymbol{k}_{10}$ takeing part in the phase transition and $[\boldsymbol{k}_{10}^{(12)}]$ means the star channel which is formed by first, second arms of $\boldsymbol{k}_{10}$ taking part in the phase transition. As shown, the structural phase transitions with volume changes of 2 times and 4 times are respectively occurred by 2-, 4-dimensional representations of substar group, while Landau symmetry phases coincide with the ones induced by 6-, 12-dimensional representations of Stokes & Hatch (1988). In Table 4, Landau symmetry phases $C_{2h}^6$, $D_4^1$, $D_{2d}^2$, $S_2^1$, $D_2^1$, $C_2^3$, $C_s^4$, $C_2^1$ and so on, are related with the star channel $[\boldsymbol{k}_{10}^{(12)}]$ and the volume change is all 4 times. In Stokes & Hatch (1988), in describing the phases related with the volume change of 4 times 6-, 12-dimensional representations were used, but, in this work, 4-dimensional representation was used.

In Khalyavin & Senos (2005), the symmetry properties in La$_{2/3}$Mg$_{1/2}$W$_{1/2}$O$_3$ (La$_4$Mg$_3$W$_3$O$_{18}$) compound, exhibiting due to distributions of the lanthanum ions and the vacancies in the [La$_{1/3}$O]$'$ and [La$_{1/3}$O]$''$ layers and the rotation of the oxygen octahedra, were studied. The compound shows interesting and useful properties such as high temperature superconducting and giant magnetoresistance phenomena, forming translational symmetry superstructures. At temperature higher than 700K, the symmetry is orthorhobic $D_{2h}^{26}$. Below this temperature, the symmetry is lowered to monoclinic $C_{2h}^3$ as a result of a successive phase transition involving an anti-phase rotation of the octahedral around the axis parallel to the Lanthanum-formed rows. By Khalyavin & Senos (2005), the rotation of the octahedral around the (001) axis relates with both the representation $T_1$ of Lifshitz star $\boldsymbol{k}_{13}$ and the representation $T_2$ of non-Lifshitz star $\boldsymbol{k}_6$, of the space group $O_h^1$, where Lifshitz star $\boldsymbol{k}_{13}$ has an arm and non-Lifshitz star $\boldsymbol{k}_6$ has 12 arms. In Table 5, the results of Khalyavin & Senos (2005) and this work are showed contrastingly.

As shown, by the result of Khalyavin & Senos (2005), the space groups $D_{2h}^{26}$, $C_{2h}^3$ and $C_{2h}^6$ of lower-symmetry phases are induced by the 13-, 16-dimensional reducible representation $T_1+T_2$, $T_9+T_1+T_2$ of the space group $O_h^1$ of the parent phase, while by the result of this work, the space group $D_{2h}^{26}$ is induced by the 5-dimensional reducible representation of the star channel group $G_{[\boldsymbol{k}_{13},\,\boldsymbol{k}_6^{(1,2,9,10)}]}$ and the space group $C_{2h}^3$, $C_{2h}^6$ is induced by the 8-dimensional reducible representation of the star channel group $G_{[\boldsymbol{k}_{13},\,\boldsymbol{k}_{13},\,\boldsymbol{k}_6^{(1,2,9,10)}]}$.

**Table 4**. The Landau symmetry phases related with Lifshitz star $k_{10}$ of space group $O_h^2$

| Stokes & Hatch (1988) | | | | | This work | | | | |
|---|---|---|---|---|---|---|---|---|---|
| $(\vec{\eta})$ | T(dim.) | $G_\alpha$ | | V'/V | [k] | $(\vec{\eta})$ | T(dim.) | $G_\alpha$ | | V'/V |
| $(\eta_1\,0\,0\,0\,0)$ | | $D_4^1$ | $D_4^5$ | | | $(\eta_1\,0)$ | | $D_4^1$ | $D_4^5$ | |
| $(\eta_1\,\eta_1\,0\,0\,0)$ | $T_1$ \| $T_2$ | $C_{4h}^3$ | $C_{4h}^4$ | | | $(\eta_1\,\eta_1)$ | $T_1$ \| $T_2$ | $C_{4h}^3$ | $C_{4h}^4$ | |
| $(\eta_1\,\eta_2\,0\,0\,0)$ | | $C_4^1$ | $C_4^3$ | | $[k_{10}^{(1)}]$ | $(\eta_1\,\eta_2)$ | | $C_4^1$ | $C_4^3$ | |
| $(\eta_1\,0\,0\,0\,0\,\eta_2\,0\,0\,0\,0)$ | | $C_{2h}^5$ | | 2 | $[k_{10}^{(2)}]$ | $(\eta_1\,0\,\eta_2\,0)$ | | $C_{2h}^5$ | | 2 |
| $(\eta_1\,\eta_1\,0\,0\,0\,\eta_2\,\eta_2\,0\,0\,0)$ | | $D_2^2$ | | | $[k_{10}^{(3)}]$ | $(\eta_1\,\eta_1\,\eta_2\,\eta_2)$ | | $D_2^2$ | | |
| $(\eta_1\,\eta_2\,0\,0\,0\,-\eta_2\,\eta_1\,0\,0\,0)$ | $T_{3+4}$ | $D_2^5$ | | | | $(\eta_1\,\eta_2\,\eta_2\,-\eta_1)$ | $T_{3+4}$ | $D_2^5$ | | |
| $(\eta_1\,\eta_2\,0\,0\,0\,\eta_3\,\eta_4\,0\,0\,0)$ | | $C_2^2$ | | | | $(\eta_1\,\eta_2\,\eta_3\,\eta_4)$ | | $C_2^2$ | | |
| $(\eta_1\,\eta_1\,\eta_1\,\eta_1\,0\,0)$ | | $C_{2h}^6$ | | | | $(\eta_1\,-\eta_1\,\eta_1\,-\eta_1)$ | | $C_{2h}^6$ | | |
| $(\eta_1\,\eta_2\,\eta_1\,\eta_2\,0\,0\,-\eta_2\,-\eta_1\,-\eta_2\,\eta_1\,0\,0)$ | | | | | | | | | | |
| $(\eta_1\,0\,\eta_1\,0\,0\,0)$ | | $D_4^1$ | | | | $(\eta_1\,0\,\eta_1\,0)$ | | $D_4^1$ | $D_4^2$ | |
| $(\eta_1\,\eta_1\,\eta_1\,\eta_1\,0\,0\,\eta_2\,\eta_2\,\eta_2\,\eta_2\,0\,0)$ | $T_1,\,T_2$ | $D_4^2$ | | | | | | | | |
| $(\eta_1\,\eta_1\,\eta_1\,-\eta_1\,0\,0)$ | (6) | $D_{2d}^2$ | | | | $(0\,\eta_1\,-\eta_1\,0)$ | | $D_{2d}^2$ | $D_{2d}^4$ | |
| $(\eta_1\,\eta_2\,\eta_2\,\eta_1\,0\,0\,-\eta_2\,\eta_1\,\eta_1\,-\eta_2\,0\,0)$ | | $D_{2d}^4$ | | | | | | | | |
| $(\eta_1\,\eta_1\,\eta_2\,\eta_2\,0\,0)$ | | $S_2^1$ | | | | $(\eta_1\,-\eta_1\,\eta_2\,\eta_2)$ | | $S_2^1$ | | |
| $(\eta_1\,\eta_2\,\eta_3\,\eta_4\,0\,0\,-\eta_2\,\eta_1\,-\eta_4\,\eta_3\,0\,0)$ | | | | | | | | | | |
| $(\eta_1\,0\,\eta_2\,0\,0\,0)$ | | $D_2^1$ | | 4 | $[k_{10}^{(12)}]$ | $(\eta_1\,0\,\eta_2\,0)$ | | $D_2^1$ | $D_2^3$ | 4 |
| $(\eta_1\,\eta_1\,\eta_2\,\eta_2\,0\,0\,\eta_3\,\eta_3\,\eta_4\,\eta_4\,0\,0)$ | | $D_2^3$ | | | $[k_{10}^{(13)}]$ | | | | | |
| $(\eta_1\,\eta_1\,\eta_1\,-\eta_2\,0\,0)$ | | $D_2^1$ | | | $[k_{10}^{(23)}]$ | $(\eta_1\,0\,0\,\eta_2)$ | $T_1$ \| $T_2$ | $D_2^1$ | $D_2^3$ | |
| $(\eta_1\,\eta_2\,\eta_3\,\eta_4\,0\,0\,-\eta_2\,\eta_1\,\eta_4\,-\eta_3\,0\,0)$ | | $D_2^3$ | | | | | (4) \| (4) | | | |
| $(\eta_1\,\eta_2\,\eta_1\,\eta_2\,0\,0)$ | | $C_2^3$ | | | | $(\eta_1\,\eta_2\,\eta_1\,\eta_2)$ | | $C_2^3$ | | |
| $(\eta_1\,\eta_2\,\eta_1\,\eta_2\,0\,0\,\eta_3\,\eta_4\,\eta_3\,\eta_4\,0\,0)$ | $T_{3+4}$ | | | | | | | | | |
| $(\eta_1\,\eta_2\,\eta_1\,-\eta_2\,0\,0)$ | (12) | $C_s^4$ | | | | $(\eta_1\,\eta_2\,-\eta_2\,\eta_1)$ | | $C_s^4$ | | |
| $(\eta_1\,\eta_2\,\eta_2\,\eta_1\,0\,0\,\eta_3\,\eta_4\,\eta_4\,\eta_3\,0\,0)$ | | | | | | | | | | |
| $(\eta_1\,\eta_1\,\eta_2\,\eta_3\,0\,0)$ | | $C_2^1$ | | | | $(\eta_1\,0\,\eta_2\,\eta_3)$ | | $C_2^1$ | | |
| $(\eta_1\,\eta_2\,\eta_3\,\eta_4\,0\,0\,-\eta_2\,\eta_1\,\eta_5\,\eta_6\,0\,0)$ | | $C_2^2$ | | | | | | $C_2^2$ | | |
| $(\eta_1\,\eta_2\,\eta_3\,\eta_4\,0\,0)$ | | $C_1^1$ | | | | $(\eta_1\,\eta_2\,\eta_3\,\eta_4)$ | | $C_1^1$ | | |
| $(\eta_1\,\eta_2\,\eta_3\,\eta_4\,0\,0\,\eta_5\,\eta_6\,\eta_7\,\eta_8\,0\,0)$ | | | | | | | | | | |
| | | | | | | $(0\,0\,\eta_1\,\eta_1)$ | | $C_{2h}^5$ | | |
| * | * | * | | | | $(\eta_1\,0\,0\,0)$ | | $D_2^2$ | | 2 |
| | | | | | | $(\eta_1\,\eta_2\,0\,0)$ | | $C_2^2$ | | |

Although the space groups of the lower-symmetry phases are obtained equally, the corresponding *L*-groups differ from entirely. Therefore, the numbers of invariants constructing the thermodynamic potentials differ from respectively, too. Thus, the different potential models are obtained, which are described by the 5-, 13-componential order parameters, respectively.

**Table 5.** The results describing symmetry changes of $La_{2/3}Mg_{1/2}W_{1/2}O_3$ ($La_4Mg_3W_3O_{18}$) compound

| star | $T$ | Khalyavin & Senos (2005) | | This work | | | |
|---|---|---|---|---|---|---|---|
| | | $\eta$ | $G_\alpha$ | Star channels | $\eta$ | $G_\alpha$ | Vectors of Bravais cell |
| $k_{13}, k_6$ | $T_1+T_2$ | (b00000-c000000) | $D_{2h}^{26}$ | $[k_{13}, k_6^{(1,2,9,10)}]$ ($[k_{13}, k_6^{(3,4,11,12)}]$, $[k_{13}, k_6^{(5,6,7,8)}]$) | $(\eta_1 00\eta_2\eta_2)$ | $D_{2h}^{26}$ | (004,200,020) |
| $k_{13}, k_{13}, k_6$ | $T_9+T_1+T_2$ | (a00bc00 000000000) | $C_{2h}^3$ | $[k_{13}, k_{13}, k_6^{(1,2,9,10)}]$ | $(0\eta_1 0\eta_2 00\eta_3\eta_3)$ | $C_{2h}^3$ | (002,402,020) |
| | | (a00b0000-c000000) | $C_{2h}^6$ | ($[k_{13}, k_{13}, k_6^{(3,4,11,12)}]$, | $(\eta_1 00\eta_2 00\eta_3\eta_3)$ | $C_{2h}^6$ | (002,022,400) |
| | | (a00b00-c000000000) | $C_{2h}^6$ | $[k_{13}, k_{13}, k_6^{(5,6,7,8)}]$) | $(00\eta_1\eta_2 00\eta_3\eta_3)$ | $C_{2h}^6$ | (400,420,002) |

As shown, if the star channel group is used in, the physical properties observed in the experiment can be studied by the simplified thermodynamic potential model. Thus, to consider the actual symmetry of system in studying the translation symmetry breaking phenomena is of key importance in overcoming difficulties generated by multi-component order parameters and the nonlinearity of potential model.

### 5. Conclusion

In the paper, we established the theory of the wave vector substar group and its representation, i.e., we defined newly the wave vector substar group and its little group related with some arms of wave vector star in reciprocal space, and showed the methods to find out those irreducible representations, so that we can use effectively the representation theory of space group in studying the physical properties related with translational symmetry breaking of crystal. We revealed that the translational symmetry breaking can be descried by 250 substars and 2811 wave vector star channel groups with respect to 80 Lifshitz stars of 230 space groups,

Through comparison with the previous research results, we showed that the theory of the wave vector substar group and its representation is a new mathematical tool which let us consider effectively the actual symmetry of crystal.

## Appendix A

**Table A1**. The wave vector channel of simple cubic lattice($\Gamma_c$)

| Star | Izyumov & Syromyatnikov (1990) | | This work | Star | Izyumov & Syromyatnikov (1990) | | This work |
|---|---|---|---|---|---|---|---|
| | Chan. | n | [$k$] | | Chan. | n | [$k$] |
| $k_{12} = 0$ | $k_{12}$ | 1 | [$k$] | | (i) | 2 | [$k_{10}^{(1)}$], [$k_{10}^{(2)}$], [$k_{10}^{(3)}$] |
| $k_{13} = \frac{1}{2}(b_1 + b_2 + b_3)$ | $k_{13}$ | 2 | | $k_{10}^{(1)} = \frac{1}{2}b_3$ | (ij) | 4 | [$k_{10}^{(12)}$], [$k_{10}^{(13)}$], [$k_{10}^{(23)}$] |
| $k_{11}^{(1)} = \frac{1}{2}(b_1+b_2)$ | (i) | 2 | [$k_{11}^{(1)}$], [$k_{11}^{(2)}$], [$k_{11}^{(3)}$] | $k_{10}^{(2)} = \frac{1}{2}b_2$ | | | |
| $k_{11}^{(2)} = \frac{1}{2}(b_1+b_3)$ | (ij) | 4 | (*) | $k_{10}^{(3)} = \frac{1}{2}b_1$ | (123) | 8 | [$k_{10}^{(123)}$] |
| $k_{11}^{(3)} = \frac{1}{2}(b_2+b_3)$ | (123) | | [$k_{11}^{(123)}$] | | | | |

Table 2. The Elements of wave vector substar point groups and small groups of simple cubic lattice($\Gamma_c$)

| Star | Substar | Point group and elements of substar group | | Point group and elements of small group | |
|---|---|---|---|---|---|
| $k_{12}$ | $k_{12}$ | $O_h$ | 1~48 | $O_h$ | 1~48 |
| $k_{13}$ | $k_{13}$ | | | | |
| $k_{10}$ | $k_{10}^{(1)}$ | $D_{4h}$ | 1~4, 13~16, 25~28, 37~40 | $D_{4h}$ | 1~4, 13~16, 25~28, 37~40 |
| | $k_{10}^{(2)}$ | | 1~4, 21~24, 25~28, 45~48 | | 1~4, 21~24, 25~28, 45~48 |
| | $k_{10}^{(3)}$ | | 1~4, 17~20, 25~28, 41~44 | | 1~4, 17~20, 25~28, 41~44 |
| | $k_{10}^{(12)}$ | | | $D_{2h}$ | 1~4, 25~28 |
| | $k_{10}^{(13)}$ | | 1~4, 21~24, 25~28, 45~48 | | |
| | $k_{10}^{(23)}$ | | 1~4, 13~16, 25~28, 37~40 | | |
| | $k_{10}^{(123)}$ | $O_h$ | 1~48 | $D_{4h}$ | 1~4, 13~16, 25~28, 37~40 |
| $k_{11}$ | $k_{11}^{(1)}$ | $D_{4h}$ | 1~4, 13~16, 25~28, 37~40 | $D_{4h}$ | 1~4, 13~16, 25~28, 37~40 |
| | $k_{11}^{(2)}$ | | 1~4, 21~24, 25~28, 45~48 | | 1~4, 21~24, 25~28, 45~48 |
| | $k_{11}^{(3)}$ | | 1~4, 17~20, 25~28, 41~44 | | 1~4, 17~20, 25~28, 41~44 |
| | $k_{11}^{(12)}$ | | | $D_{2h}$ | 1~4, 25~28 |
| | $k_{11}^{(13)}$ | | 1~4, 21~24, 25~28, 45~48 | | |
| | $k_{11}^{(23)}$ | | 1~4, 13~16, 25~28, 37~40 | | |
| | $k_{11}^{(123)}$ | $O_h$ | 1~48 | $D_{4h}$ | 1~4, 13~16, 25~28, 37~40 |

**Appendix B**

In the tables below, the numbers 1, 2, ... , 48 mean respectively the rotational elements $(h_1|000)$, $(h_2|000)$, ... , $(h_{48}|000)$ of simple cubic lattice, $O_h^1$. $T_1$, $T_2$, etc. denote the full irreducible representations of Lifshitz wave vector star channel groups of space group $O_h^1$, and $B1$, $\overline{B4}$ etc. mean the symbols of representation matrices, e. g., $B1$ denotes the matrix $\begin{pmatrix} 1 & \\ & 1 \end{pmatrix}$ and $\overline{B4}$ denotes the matrix $\begin{pmatrix} -1 & \\ & 1 \end{pmatrix}$ opposite to the matrix $B4$ $\begin{pmatrix} 1 & \\ & -1 \end{pmatrix}$.

The symbols of some representations are as follows. (where $\varepsilon = -\dfrac{1}{2} + \dfrac{\sqrt{3}}{2}$, $\varepsilon^2 = -\dfrac{1}{2} - \dfrac{\sqrt{3}}{2}$)

$$
\begin{array}{cccccccc}
B1 & B2 & B3 & B4 & B11 & B12 & B13 & B14 \\
\begin{pmatrix} 1 & \\ & 1 \end{pmatrix} & \begin{pmatrix} 1 & \\ & -1 \end{pmatrix} & \begin{pmatrix} & 1 \\ 1 & \end{pmatrix} & \begin{pmatrix} & 1 \\ -1 & \end{pmatrix} & \begin{pmatrix} \varepsilon & \\ & \varepsilon^2 \end{pmatrix} & \begin{pmatrix} \varepsilon^2 & \\ & \varepsilon \end{pmatrix} & \begin{pmatrix} & \varepsilon \\ \varepsilon^2 & \end{pmatrix} & \begin{pmatrix} & \varepsilon^2 \\ \varepsilon & \end{pmatrix}
\end{array}
$$

$$
\begin{array}{cccccccccc}
C1 & C2 & C3 & C4 & C5 & C6 & C7 & C8 & C9 & C10 \\
\begin{pmatrix} 1 & & \\ & 1 & \\ & & 1 \end{pmatrix} &
\begin{pmatrix} & 1 & \\ & & 1 \\ 1 & & \end{pmatrix} &
\begin{pmatrix} & & 1 \\ 1 & & \\ & 1 & \end{pmatrix} &
\begin{pmatrix} -1 & & \\ & 1 & \\ & & 1 \end{pmatrix} &
\begin{pmatrix} -1 & & \\ & 1 & \\ & & -1 \end{pmatrix} &
\begin{pmatrix} -1 & & \\ 1 & & \\ & & -1 \end{pmatrix} &
\begin{pmatrix} -1 & & \\ & & 1 \\ & 1 & \end{pmatrix} &
\begin{pmatrix} & 1 & \\ & & 1 \\ 1 & & \end{pmatrix} &
\begin{pmatrix} 1 & & \\ & & 1 \\ & 1 & \end{pmatrix} &
\begin{pmatrix} & & 1 \\ 1 & & \\ & 1 & \end{pmatrix}
\end{array}
$$

$$
\begin{array}{ccccccccc}
C11 & C12 & C13 & C14 & C15 & C16 & C17 & C18 & C19 \\
\begin{pmatrix} 1 & & \\ & -1 & \\ & & -1 \end{pmatrix} &
\begin{pmatrix} & 1 & \\ 1 & & \\ & & -1 \end{pmatrix} &
\begin{pmatrix} & -1 & \\ & & -1 \\ -1 & & \end{pmatrix} &
\begin{pmatrix} & & -1 \\ 1 & & \\ & -1 & \end{pmatrix} &
\begin{pmatrix} 1 & & \\ & -1 & \\ & & -1 \end{pmatrix} &
\begin{pmatrix} -1 & & \\ & -1 & \\ & & 1 \end{pmatrix} &
\begin{pmatrix} -1 & & \\ & 1 & \\ & & -1 \end{pmatrix} &
\begin{pmatrix} & -1 & \\ & & 1 \\ 1 & & \end{pmatrix} &
\begin{pmatrix} 1 & & \\ & & 1 \\ & -1 & \end{pmatrix}
\end{array}
$$

$$
\begin{array}{cccccccc}
C20 & C21 & C22 & C23 & C24 & E1 & E2 & E3 & E32 \\
\begin{pmatrix} 1 & & \\ & -1 & \\ & & 1 \end{pmatrix} &
\begin{pmatrix} & -1 & \\ 1 & & \\ & & 1 \end{pmatrix} &
\begin{pmatrix} & 1 & \\ -1 & & \\ & & 1 \end{pmatrix} &
\begin{pmatrix} & -1 & \\ 1 & & \\ & & 1 \end{pmatrix} &
\begin{pmatrix} & & 1 \\ & 1 & \\ -1 & & \end{pmatrix} &
\begin{pmatrix} B1 & \\ & B1 \end{pmatrix} &
\begin{pmatrix} & B1 \\ B1 & \end{pmatrix} &
\begin{pmatrix} B1 & \\ & B1 \end{pmatrix} &
\begin{pmatrix} & \overline{C8} \\ \overline{C8} & \end{pmatrix}
\end{array}
$$

$$
\begin{array}{cccccccc}
E39 & E40 & E81 & E85 & E86 & E87 & E88 & E131 \\
\begin{pmatrix} \overline{B3} & & \\ \overline{B3} & & \\ & & \overline{B3} \end{pmatrix} &
\begin{pmatrix} B3 & & \\ & B3 & \\ B3 & & \end{pmatrix} &
\begin{pmatrix} & \overline{B2} \\ \overline{B1} & \\ & B2 \end{pmatrix} &
\begin{pmatrix} & \overline{B1} \\ & B2 \\ \overline{B2} & \end{pmatrix} &
\begin{pmatrix} C14 & \\ & C14 \end{pmatrix} &
\begin{pmatrix} C16 & \\ & C16 \end{pmatrix} &
\begin{pmatrix} C15 & \\ & C15 \end{pmatrix} &
\begin{pmatrix} & B2 \\ \overline{B2} & \\ & \overline{B1} \end{pmatrix}
\end{array}
$$

$$
\begin{array}{cccccccc}
E132 & E133 & E134 & E251 & E252 & E253 & E254 & E255 \\
\begin{pmatrix} & \overline{B1} \\ B2 & \\ & \overline{B2} \end{pmatrix} &
\begin{pmatrix} & B2 \\ & \overline{B2} \\ \overline{B1} & \end{pmatrix} &
\begin{pmatrix} & \overline{B2} \\ & \overline{B1} \\ B2 & \end{pmatrix} &
\begin{pmatrix} B4 & \\ & \overline{B4} \\ & \overline{B3} \end{pmatrix} &
\begin{pmatrix} \overline{B4} & \\ & \overline{B4} \\ B4 & \end{pmatrix} &
\begin{pmatrix} \overline{B3} & \\ & \overline{B3} \\ & \overline{B4} \end{pmatrix} &
\begin{pmatrix} & B4 \\ B4 & \overline{B4} \\ & \overline{B3} \end{pmatrix} &
\begin{pmatrix} \overline{B4} & \\ & \overline{B3} \\ & B4 \end{pmatrix}
\end{array}
$$

$$
\begin{array}{cccc}
E256 & E257 & E258 & E259 \\
\begin{pmatrix} & \overline{B3} \\ B4 & \\ & \overline{B4} \end{pmatrix} &
\begin{pmatrix} & \overline{C21} \\ \overline{C21} & \end{pmatrix} &
\begin{pmatrix} & C13 \\ C13 & \end{pmatrix} &
\begin{pmatrix} & \overline{C24} \\ \overline{C24} & \end{pmatrix}
\end{array}
$$

| $k_{10}[1], k_{11}[1]$ | | 1 | 2 | 3 | 4 | 13 | 14 | 15 | 16 | 25 | 26 | 27 | 28 | 37 | 38 | 39 | 40 |
|---|---|---|---|---|---|---|---|---|---|---|---|---|---|---|---|---|---|
| [2], | [2] | 1 | 4 | 2 | 3 | 21 | 24 | 22 | 23 | 25 | 28 | 26 | 27 | 45 | 48 | 46 | 47 |
| [3], | [3] | 1 | 3 | 4 | 2 | 17 | 19 | 20 | 18 | 25 | 27 | 28 | 26 | 41 | 43 | 44 | 42 |
| | $T_1$ | 1 | 1 | 1 | 1 | 1 | 1 | 1 | 1 | 1 | 1 | 1 | 1 | 1 | 1 | 1 | 1 |
| | $T_2$ | 1 | 1 | 1 | 1 | 1 | 1 | 1 | 1 | -1 | -1 | -1 | -1 | -1 | -1 | -1 | -1 |
| | $T_3$ | 1 | -1 | -1 | 1 | -1 | 1 | 1 | -1 | 1 | -1 | -1 | 1 | -1 | 1 | 1 | -1 |
| | $T_4$ | 1 | -1 | -1 | 1 | -1 | 1 | 1 | -1 | -1 | 1 | 1 | -1 | 1 | -1 | -1 | 1 |
| | $T_5$ | 1 | 1 | 1 | 1 | -1 | -1 | -1 | -1 | 1 | 1 | 1 | 1 | -1 | -1 | -1 | -1 |
| | $T_6$ | 1 | 1 | 1 | 1 | -1 | -1 | -1 | -1 | -1 | -1 | -1 | -1 | 1 | 1 | 1 | 1 |
| | $T_7$ | 1 | -1 | -1 | 1 | 1 | -1 | -1 | 1 | 1 | -1 | -1 | 1 | 1 | -1 | -1 | 1 |
| | $T_8$ | 1 | -1 | -1 | 1 | 1 | -1 | -1 | 1 | -1 | 1 | 1 | -1 | -1 | 1 | 1 | -1 |
| | $T_9$ | $B1$ | $B2$ | $\overline{B2}$ | $\overline{B1}$ | $B3$ | $B4$ | $\overline{B4}$ | $\overline{B3}$ | $B1$ | $B2$ | $\overline{B2}$ | $\overline{B1}$ | $B3$ | $B4$ | $\overline{B4}$ | $\overline{B3}$ |
| | $T_{10}$ | $B1$ | $B2$ | $\overline{B2}$ | $\overline{B1}$ | $B3$ | $B4$ | $\overline{B4}$ | $\overline{B3}$ | $\overline{B1}$ | $\overline{B2}$ | $B2$ | $B1$ | $\overline{B3}$ | $\overline{B4}$ | $B4$ | $B3$ |

| $k_{10}[12]$ | | 1 | 2 | 3 | 4 | 17 | 18 | 19 | 20 | 25 | 26 | 27 | 28 | 41 | 42 | 43 | 44 |
|---|---|---|---|---|---|---|---|---|---|---|---|---|---|---|---|---|---|
| [13] | | 1 | 3 | 4 | 2 | 21 | 23 | 24 | 22 | 25 | 27 | 28 | 26 | 45 | 47 | 48 | 46 |
| [23] | | 1 | 4 | 2 | 3 | 13 | 16 | 14 | 15 | 25 | 28 | 26 | 27 | 37 | 40 | 38 | 39 |
| | $T_1$ | $B1$ | $B1$ | $B1$ | $B1$ | $B3$ | $B3$ | $B3$ | $B3$ | $B1$ | $B1$ | $B1$ | $B1$ | $B3$ | $B3$ | $B3$ | $B3$ |
| | $T_2$ | $B1$ | $B1$ | $B1$ | $B1$ | $\overline{B3}$ | $\overline{B3}$ | $\overline{B3}$ | $\overline{B3}$ | $\overline{B1}$ | $\overline{B1}$ | $\overline{B1}$ | $\overline{B1}$ | $B3$ | $B3$ | $B3$ | $B3$ |
| | $T_3$ | $B1$ | $B1$ | $\overline{B1}$ | $\overline{B1}$ | $B3$ | $B3$ | $\overline{B3}$ | $\overline{B3}$ | $B1$ | $B1$ | $\overline{B1}$ | $\overline{B1}$ | $B3$ | $B3$ | $\overline{B3}$ | $\overline{B3}$ |
| | $T_4$ | $B1$ | $B1$ | $\overline{B1}$ | $\overline{B1}$ | $B3$ | $B3$ | $B3$ | $B3$ | $\overline{B1}$ | $\overline{B1}$ | $B1$ | $B1$ | $B3$ | $B3$ | $\overline{B3}$ | $\overline{B3}$ |
| | $T_5$ | $B1$ | $\overline{B1}$ | $B2$ | $\overline{B2}$ | $B3$ | $\overline{B3}$ | $B4$ | $\overline{B4}$ | $B1$ | $\overline{B1}$ | $B2$ | $\overline{B2}$ | $B3$ | $\overline{B3}$ | $B4$ | $\overline{B4}$ |
| | $T_6$ | $B1$ | $\overline{B1}$ | $B2$ | $\overline{B2}$ | $\overline{B3}$ | $B3$ | $\overline{B4}$ | $B4$ | $\overline{B1}$ | $B1$ | $\overline{B2}$ | $B2$ | $B3$ | $\overline{B3}$ | $B4$ | $\overline{B4}$ |
| | $T_7$ | $B1$ | $\overline{B1}$ | $\overline{B2}$ | $B2$ | $B3$ | $\overline{B3}$ | $\overline{B4}$ | $B4$ | $B1$ | $\overline{B1}$ | $\overline{B2}$ | $B2$ | $B3$ | $\overline{B3}$ | $\overline{B4}$ | $B4$ |
| | $T_8$ | $B1$ | $\overline{B1}$ | $\overline{B2}$ | $B2$ | $\overline{B3}$ | $B3$ | $B4$ | $\overline{B4}$ | $\overline{B1}$ | $B1$ | $B2$ | $\overline{B2}$ | $B3$ | $\overline{B3}$ | $\overline{B4}$ | $B4$ |
| | $T_9$ | $B1$ | $B1$ | $B1$ | $B1$ | $B3$ | $B3$ | $B3$ | $B3$ | $B1$ | $B1$ | $B1$ | $B1$ | $B3$ | $B3$ | $B3$ | $B3$ |
| | $T_{10}$ | $B1$ | $B1$ | $B1$ | $B1$ | $\overline{B3}$ | $\overline{B3}$ | $\overline{B3}$ | $\overline{B3}$ | $\overline{B1}$ | $\overline{B1}$ | $\overline{B1}$ | $\overline{B1}$ | $B3$ | $B3$ | $B3$ | $B3$ |

$k_{10}[123], k_{11}[123]$

| | 1 | 2 | 3 | 4 | 5 | 6 | 7 | 8 | 9 | 10 | 11 | 12 | 13 | 14 | 15 | 16 |
|---|---|---|---|---|---|---|---|---|---|---|---|---|---|---|---|---|
| $T_1$ | $C1$ | $C1$ | $C1$ | $C1$ | $C2$ | $C2$ | $C2$ | $C2$ | $C3$ | $C3$ | $C3$ | $C3$ | $C9$ | $C9$ | $C9$ | $C9$ |
| $T_2$ | $C1$ | $C1$ | $C1$ | $C1$ | $C2$ | $C2$ | $C2$ | $C2$ | $C3$ | $C3$ | $C3$ | $C3$ | $C9$ | $C9$ | $C9$ | $C9$ |
| $T_3$ | $C1$ | $C15$ | $C16$ | $C14$ | $C2$ | $C17$ | $C6$ | $\overline{C4}$ | $C3$ | $C5$ | $C18$ | $\overline{C7}$ | $\overline{C9}$ | $C19$ | $C20$ | $\overline{C11}$ |
| $T_4$ | $C1$ | $C15$ | $C16$ | $C14$ | $C2$ | $C17$ | $C6$ | $\overline{C4}$ | $C3$ | $C5$ | $C18$ | $\overline{C7}$ | $\overline{C9}$ | $C19$ | $C20$ | $\overline{C11}$ |
| $T_5$ | $C1$ | $C1$ | $C1$ | $C1$ | $C2$ | $C2$ | $C2$ | $C2$ | $C3$ | $C3$ | $C3$ | $C3$ | $\overline{C9}$ | $\overline{C9}$ | $\overline{C9}$ | $\overline{C9}$ |
| $T_6$ | $C1$ | $C1$ | $C1$ | $C1$ | $C2$ | $C2$ | $C2$ | $C2$ | $C3$ | $C3$ | $C3$ | $C3$ | $\overline{C9}$ | $\overline{C9}$ | $\overline{C9}$ | $\overline{C9}$ |
| $T_7$ | $C1$ | $C15$ | $C16$ | $C14$ | $C2$ | $C17$ | $C6$ | $\overline{C4}$ | $C3$ | $C5$ | $C18$ | $\overline{C7}$ | $C9$ | $\overline{C19}$ | $\overline{C20}$ | $C11$ |
| $T_8$ | $C1$ | $C15$ | $C16$ | $C14$ | $C2$ | $C17$ | $C6$ | $\overline{C4}$ | $C3$ | $C5$ | $C18$ | $\overline{C7}$ | $C9$ | $\overline{C19}$ | $\overline{C20}$ | $C11$ |
| $T_9$ | $E1$ | $E86$ | $E87$ | $E88$ | $E2$ | $E131$ | $E81$ | $E132$ | $E3$ | $E133$ | $E134$ | $E85$ | $E40$ | $E251$ | $E252$ | $E253$ |
| $T_{10}$ | $E1$ | $E86$ | $E87$ | $E88$ | $E2$ | $E131$ | $E81$ | $E132$ | $E3$ | $E133$ | $E134$ | $E85$ | $E40$ | $E251$ | $E252$ | $E253$ |

$k_{10}[123], k_{11}[123]$ (continue)

|   | 17 | 18 | 19 | 20 | 21 | 22 | 23 | 24 | 25 | 26 | 27 | 28 | 29 | 30 | 31 | 32 |
|---|---|---|---|---|---|---|---|---|---|---|---|---|---|---|---|---|
| $T_1$ | C10 | C10 | C10 | C10 | C8 | C8 | C8 | C8 | C1 | C1 | C1 | C1 | C2 | C2 | C2 | C2 |
| $T_2$ | C10 | C10 | C10 | C10 | C8 | C8 | C8 | C8 | $\overline{C1}$ | $\overline{C1}$ | $\overline{C1}$ | $\overline{C1}$ | $\overline{C2}$ | $\overline{C2}$ | $\overline{C2}$ | $\overline{C2}$ |
| $T_3$ | $\overline{C10}$ | C12 | C22 | C23 | $\overline{C8}$ | C24 | $\overline{C13}$ | C21 | C1 | C15 | C16 | C14 | C2 | C17 | C6 | $\overline{C4}$ |
| $T_4$ | $\overline{C10}$ | C12 | C22 | C23 | $\overline{C8}$ | C24 | $\overline{C13}$ | C21 | $\overline{C1}$ | $\overline{C15}$ | $\overline{C16}$ | $\overline{C14}$ | $\overline{C2}$ | $\overline{C17}$ | $\overline{C6}$ | C4 |
| $T_5$ | $\overline{C10}$ | $\overline{C10}$ | $\overline{C10}$ | $\overline{C10}$ | $\overline{C8}$ | $\overline{C8}$ | $\overline{C8}$ | $\overline{C8}$ | C1 | C1 | C1 | C1 | C2 | C2 | C2 | C2 |
| $T_6$ | $\overline{C10}$ | $\overline{C10}$ | $\overline{C10}$ | $\overline{C10}$ | $\overline{C8}$ | $\overline{C8}$ | $\overline{C8}$ | $\overline{C8}$ | $\overline{C1}$ | $\overline{C1}$ | $\overline{C1}$ | $\overline{C1}$ | $\overline{C2}$ | $\overline{C2}$ | $\overline{C2}$ | $\overline{C2}$ |
| $T_7$ | C10 | $\overline{C12}$ | $\overline{C22}$ | $\overline{C23}$ | C8 | $\overline{C24}$ | C13 | $\overline{C21}$ | C1 | C15 | C16 | C14 | C2 | C17 | C6 | $\overline{C4}$ |
| $T_8$ | C10 | $\overline{C12}$ | $\overline{C22}$ | $\overline{C23}$ | C8 | $\overline{C24}$ | C13 | $\overline{C21}$ | $\overline{C1}$ | $\overline{C15}$ | $\overline{C16}$ | $\overline{C14}$ | $\overline{C2}$ | $\overline{C17}$ | $\overline{C6}$ | C4 |
| $T_9$ | $\overline{E39}$ | E254 | E255 | E256 | $\overline{E32}$ | E257 | E258 | E259 | E1 | E86 | E87 | E88 | E2 | E131 | E81 | E132 |
| $T_{10}$ | $\overline{E39}$ | E254 | E255 | E256 | $\overline{E32}$ | E257 | E258 | E259 | $\overline{E1}$ | $\overline{E86}$ | $\overline{E87}$ | $\overline{E88}$ | $\overline{E2}$ | $\overline{E131}$ | $\overline{E81}$ | $\overline{E132}$ |

$k_{10}[123], k_{11}[123]$ (continue)

|   | 33 | 34 | 35 | 36 | 37 | 38 | 39 | 40 | 41 | 42 | 43 | 44 | 45 | 46 | 47 | 48 |
|---|---|---|---|---|---|---|---|---|---|---|---|---|---|---|---|---|
| $T_1$ | C3 | C3 | C3 | C3 | C9 | C9 | C9 | C9 | C10 | C10 | C10 | C10 | C8 | C8 | C8 | C8 |
| $T_2$ | $\overline{C3}$ | $\overline{C3}$ | $\overline{C3}$ | $\overline{C3}$ | $\overline{C9}$ | $\overline{C9}$ | $\overline{C9}$ | $\overline{C9}$ | $\overline{C10}$ | $\overline{C10}$ | $\overline{C10}$ | $\overline{C10}$ | $\overline{C8}$ | $\overline{C8}$ | $\overline{C8}$ | $\overline{C8}$ |
| $T_3$ | C3 | C5 | C18 | $\overline{C7}$ | $\overline{C9}$ | C19 | C20 | $\overline{C11}$ | $\overline{C10}$ | C12 | C22 | C23 | $\overline{C8}$ | C24 | $\overline{C13}$ | C21 |
| $T_4$ | $\overline{C3}$ | $\overline{C5}$ | $\overline{C18}$ | C7 | C9 | $\overline{C19}$ | $\overline{C20}$ | C11 | C10 | $\overline{C12}$ | $\overline{C22}$ | $\overline{C23}$ | C8 | $\overline{C24}$ | C13 | $\overline{C21}$ |
| $T_5$ | C3 | C3 | C3 | C3 | $\overline{C9}$ | $\overline{C9}$ | $\overline{C9}$ | $\overline{C9}$ | $\overline{C10}$ | $\overline{C10}$ | $\overline{C10}$ | $\overline{C10}$ | $\overline{C8}$ | $\overline{C8}$ | $\overline{C8}$ | $\overline{C8}$ |
| $T_6$ | $\overline{C3}$ | $\overline{C3}$ | $\overline{C3}$ | $\overline{C3}$ | C9 | C9 | C9 | C9 | C10 | C10 | C10 | C10 | C8 | C8 | C8 | C8 |
| $T_7$ | C3 | C5 | C18 | $\overline{C7}$ | C9 | $\overline{C19}$ | $\overline{C20}$ | C11 | C10 | $\overline{C12}$ | $\overline{C22}$ | $\overline{C23}$ | C8 | $\overline{C24}$ | C13 | $\overline{C21}$ |
| $T_8$ | $\overline{C3}$ | $\overline{C5}$ | $\overline{C18}$ | C7 | $\overline{C9}$ | C19 | C20 | $\overline{C11}$ | $\overline{C10}$ | C12 | C22 | C23 | $\overline{C8}$ | C24 | $\overline{C13}$ | C21 |
| $T_9$ | E3 | E133 | E134 | E85 | E40 | E251 | E252 | E253 | $\overline{E39}$ | E254 | E255 | E256 | $\overline{E32}$ | E257 | E258 | E259 |
| $T_{10}$ | $\overline{E3}$ | $\overline{E133}$ | $\overline{E134}$ | $\overline{E85}$ | $\overline{E40}$ | $\overline{E251}$ | $\overline{E252}$ | $\overline{E253}$ | E39 | $\overline{E254}$ | $\overline{E255}$ | $\overline{E256}$ | E32 | $\overline{E257}$ | $\overline{E258}$ | $\overline{E259}$ |

$[k_{12}], [k_{13}]$

|   | 1 | 2 | 3 | 4 | 5 | 6 | 7 | 8 | 9 | 10 | 11 | 12 | 13 | 14 | 15 | 16 |
|---|---|---|---|---|---|---|---|---|---|---|---|---|---|---|---|---|
| $T_1$ | 1 | 1 | 1 | 1 | 1 | 1 | 1 | 1 | 1 | 1 | 1 | 1 | 1 | 1 | 1 | 1 |
| $T_2$ | 1 | 1 | 1 | 1 | 1 | 1 | 1 | 1 | 1 | 1 | 1 | 1 | 1 | 1 | 1 | 1 |
| $T_3$ | 1 | 1 | 1 | 1 | 1 | 1 | 1 | 1 | 1 | 1 | 1 | 1 | -1 | -1 | -1 | -1 |
| $T_4$ | 1 | 1 | 1 | 1 | 1 | 1 | 1 | 1 | 1 | 1 | 1 | 1 | -1 | -1 | -1 | -1 |
| $T_5$ | B1 | B1 | B1 | B1 | B11 | B11 | B11 | B11 | B12 | B12 | B12 | B12 | B3 | B3 | B3 | B3 |
| $T_6$ | B1 | B1 | B1 | B1 | B11 | B11 | B11 | B11 | B12 | B12 | B12 | B12 | B3 | B3 | B3 | B3 |
| $T_7$ | C1 | C14 | C16 | C15 | C3 | $\overline{C7}$ | C18 | C5 | C2 | $\overline{C4}$ | C6 | C17 | C10 | $\overline{C23}$ | $\overline{C22}$ | $\overline{C12}$ |
| $T_8$ | C1 | C14 | C16 | C15 | C3 | $\overline{C7}$ | C18 | C5 | C2 | $\overline{C4}$ | C6 | C17 | C10 | $\overline{C23}$ | $\overline{C22}$ | $\overline{C12}$ |
| $T_9$ | C1 | C14 | C16 | C15 | C3 | $\overline{C7}$ | C18 | C5 | C2 | $\overline{C4}$ | C6 | C17 | $\overline{C10}$ | C23 | C22 | C12 |
| $T_{10}$ | C1 | C14 | C16 | C15 | C3 | $\overline{C7}$ | C18 | C5 | C2 | $\overline{C4}$ | C6 | C17 | $\overline{C10}$ | C23 | C22 | C12 |

$[k_{12}]$ , $[k_{13}]$ (continue)

|   | 17 | 18 | 19 | 20 | 21 | 22 | 23 | 24 | 25 | 26 | 27 | 28 | 29 | 30 | 31 | 32 |
|---|---|---|---|---|---|---|---|---|---|---|---|---|---|---|---|---|
| $T_1$ | 1 | 1 | 1 | 1 | 1 | 1 | 1 | 1 | 1 | 1 | 1 | 1 | 1 | 1 | 1 | 1 |
| $T_2$ | 1 | 1 | 1 | 1 | 1 | 1 | 1 | 1 | -1 | -1 | -1 | -1 | -1 | -1 | -1 | -1 |
| $T_3$ | -1 | -1 | -1 | -1 | -1 | -1 | -1 | -1 | 1 | 1 | 1 | 1 | 1 | 1 | 1 | 1 |
| $T_4$ | -1 | -1 | -1 | -1 | -1 | -1 | -1 | -1 | -1 | -1 | -1 | -1 | -1 | -1 | -1 | -1 |
| $T_5$ | $B13$ | $B13$ | $B13$ | $B13$ | $B14$ | $B14$ | $B14$ | $B14$ | $B1$ | $B1$ | $B1$ | $B1$ | $B11$ | $B11$ | $B11$ | $B11$ |
| $T_6$ | $B13$ | $B13$ | $B13$ | $B13$ | $B14$ | $B14$ | $B14$ | $B14$ | $\overline{B1}$ | $\overline{B1}$ | $\overline{B1}$ | $\overline{B1}$ | $\overline{B11}$ | $\overline{B11}$ | $\overline{B11}$ | $\overline{B11}$ |
| $T_7$ | $C9$ | $C11$ | $\overline{C20}$ | $\overline{C19}$ | $C8$ | $\overline{C21}$ | $C13$ | $\overline{C24}$ | $C1$ | $C14$ | $C16$ | $C15$ | $C3$ | $\overline{C7}$ | $C18$ | $C5$ |
| $T_8$ | $C9$ | $C11$ | $\overline{C20}$ | $\overline{C19}$ | $C8$ | $\overline{C21}$ | $C13$ | $\overline{C24}$ | $\overline{C1}$ | $\overline{C14}$ | $\overline{C16}$ | $\overline{C15}$ | $\overline{C3}$ | $C7$ | $\overline{C18}$ | $\overline{C5}$ |
| $T_9$ | $\overline{C9}$ | $\overline{C11}$ | $C20$ | $C19$ | $\overline{C8}$ | $C21$ | $\overline{C13}$ | $C24$ | $C1$ | $C14$ | $C16$ | $C15$ | $C3$ | $\overline{C7}$ | $C18$ | $C5$ |
| $T_{10}$ | $\overline{C9}$ | $\overline{C11}$ | $C20$ | $C19$ | $\overline{C8}$ | $C21$ | $\overline{C13}$ | $C24$ | $\overline{C1}$ | $\overline{C14}$ | $\overline{C16}$ | $\overline{C15}$ | $\overline{C3}$ | $C7$ | $\overline{C18}$ | $\overline{C5}$ |

$[k_{12}]$ , $[k_{13}]$ (continue)

|   | 33 | 34 | 35 | 36 | 37 | 38 | 39 | 40 | 41 | 42 | 43 | 44 | 45 | 46 | 47 | 48 |
|---|---|---|---|---|---|---|---|---|---|---|---|---|---|---|---|---|
| $T_1$ | 1 | 1 | 1 | 1 | 1 | 1 | 1 | 1 | 1 | 1 | 1 | 1 | 1 | 1 | 1 | 1 |
| $T_2$ | -1 | -1 | -1 | -1 | -1 | -1 | -1 | -1 | -1 | -1 | -1 | -1 | -1 | -1 | -1 | -1 |
| $T_3$ | 1 | 1 | 1 | 1 | -1 | -1 | -1 | -1 | -1 | -1 | -1 | -1 | -1 | -1 | -1 | -1 |
| $T_4$ | -1 | -1 | -1 | -1 | 1 | 1 | 1 | 1 | 1 | 1 | 1 | 1 | 1 | 1 | 1 | 1 |
| $T_5$ | $B12$ | $B12$ | $B12$ | $B12$ | $B3$ | $B3$ | $B3$ | $B3$ | $B13$ | $B13$ | $B13$ | $B13$ | $B14$ | $B14$ | $B14$ | $B14$ |
| $T_6$ | $\overline{B12}$ | $\overline{B12}$ | $\overline{B12}$ | $\overline{B12}$ | $\overline{B3}$ | $\overline{B3}$ | $\overline{B3}$ | $\overline{B3}$ | $\overline{B13}$ | $\overline{B13}$ | $\overline{B13}$ | $\overline{B13}$ | $\overline{B14}$ | $\overline{B14}$ | $\overline{B14}$ | $\overline{B14}$ |
| $T_7$ | $C2$ | $\overline{C4}$ | $C6$ | $C17$ | $C10$ | $\overline{C23}$ | $\overline{C22}$ | $\overline{C12}$ | $C9$ | $C11$ | $\overline{C20}$ | $\overline{C19}$ | $C8$ | $\overline{C21}$ | $C13$ | $\overline{C24}$ |
| $T_8$ | $\overline{C2}$ | $C4$ | $\overline{C6}$ | $\overline{C17}$ | $\overline{C10}$ | $C23$ | $C22$ | $C12$ | $\overline{C9}$ | $\overline{C11}$ | $C20$ | $C19$ | $\overline{C8}$ | $C21$ | $\overline{C13}$ | $C24$ |
| $T_9$ | $C2$ | $\overline{C4}$ | $C6$ | $C17$ | $\overline{C10}$ | $C23$ | $C22$ | $C12$ | $\overline{C9}$ | $\overline{C11}$ | $C20$ | $C19$ | $\overline{C8}$ | $C21$ | $\overline{C13}$ | $C24$ |
| $T_{10}$ | $\overline{C2}$ | $C4$ | $\overline{C6}$ | $\overline{C17}$ | $C10$ | $\overline{C23}$ | $\overline{C22}$ | $\overline{C12}$ | $C9$ | $C11$ | $\overline{C20}$ | $\overline{C19}$ | $C8$ | $\overline{C21}$ | $C13$ | $\overline{C24}$ |